\documentclass[10pt,twocolumn]{article}
\usepackage{times}
\usepackage{cvpr}
\usepackage{graphicx}
\usepackage{amsmath}
\usepackage{amssymb}
\usepackage{booktabs} % professional-quality tables
\usepackage{enumitem}
\usepackage{textcomp}

\usepackage[square,numbers,sort,compress]{natbib}

\usepackage[pagebackref=true,breaklinks=true,colorlinks,bookmarks=false]{hyperref}

\cvprfinalcopy % *** Uncomment this line for the final submission

\def\cvprPaperID{0789} % *** Enter the CVPR Paper ID here

\ifcvprfinal\fi

\usepackage{overpic}
\usepackage{multirow} %< used in tables
\usepackage{diagbox}
\usepackage[T1]{fontenc} %< allows quotes
\usepackage[style=american]{csquotes} %< defines enquote
\usepackage{mathtools}
\usepackage{color}
\definecolor{turquoise}{cmyk}{0.65,0,0.1,0.3}
\definecolor{purple}{rgb}{0.65,0,0.65}
\definecolor{dark_green}{rgb}{0, 0.5, 0}
\definecolor{orange}{rgb}{0.8, 0.6, 0.2}
\definecolor{red}{rgb}{0.8, 0.2, 0.2}
\definecolor{blueish}{rgb}{0.0, 0.7, 1}
\definecolor{light_gray}{rgb}{0.7, 0.7, .7}
\definecolor{pink}{rgb}{1, 0, 1}
\definecolor{dark_red}{rgb}{0.5, 0, 0}

\newcommand{\hide}[1]{{}} %< discards

\renewcommand{\sc}[1]

\newcommand{\Figure}[1]{Figure~\ref{fig:#1}}

\newcommand{\eq}[1]{(\ref{eq:#1})}

\usepackage{currfile}

\usepackage{blindtext}

\usepackage{parskip}
\renewcommand{\paragraph}[1]{\vspace{\parskip}\textbf{#1}.~}

\DeclareMathAlphabet\mathbfcal{OMS}{cmsy}{b}{n}

\newcommand{\V}{\mathbf{V}}

\newcommand{\point}{\mathbf{x}}
\newcommand{\vertex}{\mathbf{v}}
\newcommand{\T}{\mathbf{B}}
\newcommand{\pose}{\boldsymbol{\theta}}

\newcommand{\poses}{\boldsymbol{\Theta}}

\newcommand{\weights}{\mathbf{w}} %< LBS weights
\newcommand{\R}{\mathbb{R}}
\newcommand{\x}{\mathbf{x}}
\newcommand{\object}{\mathcal{O}}

\newcommand{\given}{|}
\newcommand{\expect}[2]{\mathbb{E}_{#1}\left[ #2 \right]}

\newcommand{\defo}{\tilde}
\newcommand{\rest}{\bar}

\newcommand{\loss}{\mathcal{L}}

\newcommand{\crossentropy}[1]{\text{CE}\!\left( #1 \right)}

\newcommand{\back}{\mathcal{R}}
\newcommand{\fwd}{\mathcal{F}}

\begin{document}
\title{NiLBS: Neural Inverse Linear Blend Skinning}
\author{Timothy Jeruzalski$^{*+}$ \quad David I.W. Levin$^{+}$ \quad Alec Jacobson$^{+}$ \quad \\ Paul Lalonde$^{*}$ \quad Mohammad Norouzi$^{*}$ \quad Andrea Tagliasacchi$^{*}$ 
\\[1em]
$^{*}$Google Research \quad $^{+}$University of Toronto
}
\maketitle

\begin{abstract}
In this technical report, we investigate efficient representations of articulated objects (e.g.~human bodies), which is an important problem in computer vision and graphics.
To deform articulated geometry, existing approaches represent objects as meshes and deform them using ``skinning'' techniques.
The skinning operation allows a wide range of deformations to be achieved with a small number of control parameters.
This paper introduces a method to invert the deformations undergone via traditional skinning techniques via a neural network parameterized by pose.
The ability to invert these deformations allows values (e.g., distance function, signed distance function, occupancy) to be pre-computed at rest pose, and then efficiently queried when the character is deformed.
We leave empirical evaluation of our approach to future work.
\end{abstract}

\section{Linear Blend Skinning (LBS)}
Linear blend skinning is widely used in computer games and other real-time applications to deform a character's skin following the motion of an underlying abstract skeleton \cite{skinningcourse}.
Given a deformable surface mesh model whose vertices in homogeneous coordinates are $\bar\V{=}\{\bar\vertex_n {\in} \R^4 \}$\footnote{We represent deformations in 2D in the same way, and assume $z{=}0$.}.
Let us define the LBS deformation process as:
\begin{equation}
\tilde\vertex_n = \left[ \sum_b w_{bn} \T_{b} \right] \bar\vertex_n
\label{eq:lbs}
\end{equation}
where the collection of $B$ homogeneous transformations $\{\T_b {\in} \R^{4 \times 4} \}$ are blended by the weights $w_{bn}$.
These weights are typically \textit{painted} by a digital artist on the surface of the model, although automated algorithms exist for cases when less control and quality is needed~\cite{autoskin}.
Each of these transformation matrices are typically factored into the product of two transformations $\T_b {=} \tilde\T_b \bar\T^{-1}_b$.
In what follows, we will use $\bar\cdot$ to denote quantities in the \textit{rest} coordinate frame, and $\tilde\cdot$ in a \textit{deformed} coordinate frame.
The homogeneous transformations $\bar\T^{-1}_b$ are expressed in rest pose and kept fixed throughout deformation, while $\tilde\T_b$ are typically computed as a function of the \textit{pose} degrees of freedom $\pose$ -- we can abstract this process via the pose function $\{\tilde\T_b\} = \text{pose}(\pose)$.
Given a vertex in rest pose $\bar\vertex \in \bar\V$, the transformation:
\begin{equation}
\tilde\vertex_b = \tilde\T_b \bar\T^{-1}_b \bar\vertex
\label{eq:bonemap}
\end{equation}
first \textit{encodes} the point in the (local, rest) $b$-th coordinate frame as $\bar\vertex_b {=} \bar\T^{-1}_b \bar\vertex$, and then \textit{decodes} it in the $b$-th (global, posed) coordinate frame as $\tilde\vertex_b = \tilde\T_b \bar\vertex_b$.
Finally, once we represent the weights $\weights_n{=}\{\weights_{*n} \}$, as well as the transformations $\T(\theta){=}\{\T_b\}$ in tensor form, we can summarize the skinning of a single vertex as:
\begin{equation}
\tilde\vertex_n = \left[ \weights_n \T(\theta) \right] \bar\vertex_n, 
\label{eq:lbsvec}
\end{equation}
where  $\T(\theta)$ has dimensionality $B{\times}4{\times}4$, $\weights_n$ has dimensionality $1{\times}B$, and we highlight the dependency of $\T$ on the pose parameters $\pose$.
Further, $\weights_n$ are typically normalized so that they are positive and sum to one.
% $\|\weights_n \|_1{=}1$, and $\weights_n[b]{>}0 \:\:\forall{b}$.

\begin{figure}[t]
\centering
\begin{overpic} 
[width=\linewidth]
% [width=\linewidth,grid,tics=10]
{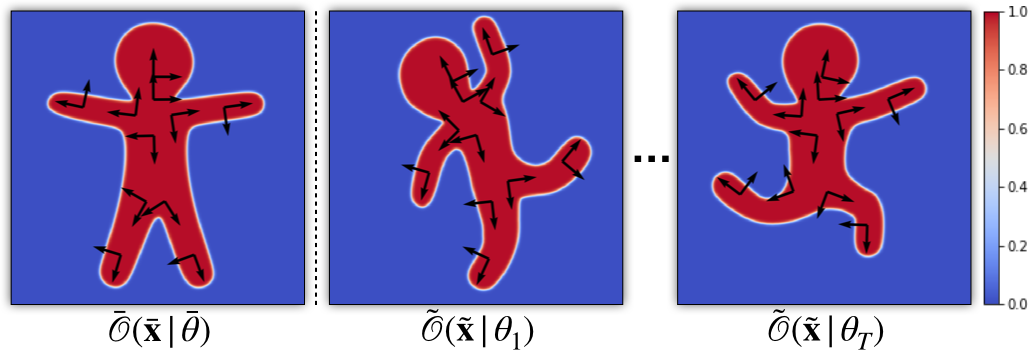}
\end{overpic}
\caption{
\textbf{Notation} --
(left) The ground truth occupancy $\rest\object(\bar\x |\rest\pose)$ in the rest frame and the $\{\rest\T_b\}_{b=1}^B$transformation frames.
(right) The training data consisting of $T$ frames of an animation with pose parameters~$\{\pose_t\}$, and corresponding occupancy~$\{\defo\object(\x \given \pose_t)\}$.
}
\label{fig:notation}
\end{figure}

\begin{figure}[ht]
\centering
\begin{overpic} 
[width=\linewidth]
% [width=\linewidth,grid,tics=10]
{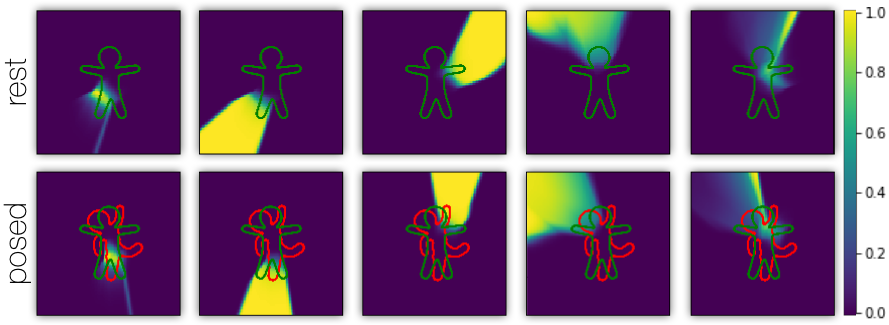}
% \myfigurename{}
\end{overpic}
\caption{
\textbf{Weights -- }
We visualize a few channels of $\mathcal{W}_\omega$.
(top)~The weights in rest pose are a volumetric version of traditional skinning weights.
(bottom)~The pose-conditional weights functions. When the weight is one, the portion of space is entirely mapped to a particular bone. 
}
\label{fig:weights}
\end{figure}

\section{Neural inverse LBS (NiLBS)}
We seek a method to query the occupancy function of a \textit{posed} character. Occupancy evaluates to $0$ outside and $1$ inside; see~\Figure{notation}.
We employ the pre-computed rest pose occupancy as a \textit{cache}.
This cache can be used to query the occupancy in a desired deformed pose, as far as a suitable \textit{mapping} function exists.

\subsection{Learnable forward mapping}
We can generalize the LBS equation~\eq{lbsvec} so as to express the deformation of \textit{arbitrary} points~$\rest\point$ in rest space.
This is different from traditional LBS, where the weight is defined \textit{only} at the vertices of the surface mesh.
This can be achieved by replacing the artist-painted weights $w_{bn}$ by a neural function $\mathcal{W}_\omega(\rest\point){:}\R^3 {\rightarrow} \R^{1{\times}B}$ with parameters $\omega$:
\begin{align}
\defo\point &= \fwd_\omega(\rest\point \given \pose) = 
\left[ \mathcal{W}_\omega(\rest\point) \T(\pose) \right] \rest\point
\label{eq:forward}
\end{align}
where for $\mathcal{W}_\omega$ we use an MLP with $B$ output channels, and a softmax output layer to mimic the typical LBS setup, i.e.~weights bounded to [0,1] and with unit sum.

\subsection{Learnable inverse mapping}
One would be tempted to invert the bracketed term in \eq{forward} to obtain an inverse mapping as: 
\begin{equation}
\rest\point = \back_\omega(\defo\point \given \pose) \stackrel{?}{=}
\left[ \underline{\mathcal{W}_\omega(\rest\point)} \T(\pose) \right]^{-1} \defo\point
\label{eq:forward_wrong}
\end{equation}
In the above equation, note the term $\mathcal{W}_\omega(\rest\point)$ cannot be computed, as it is parametric in the unknown~$\rest\point$.
We address this by replacing $\mathcal{W}_\omega(\rest\point)$ with $\mathcal{W}_\omega(\defo\point \given \pose)$; see~\Figure{weights}.
Hence, the network $\mathcal{W}_\omega$ is provided with all the available test time information -- query $\defo\point$ and pose $\pose$:
\begin{equation}
\rest\point = \back_\omega(\defo\point \given \pose) = 
\left[ \mathcal{W}_\omega(\defo\point \given \pose) \T(\pose) \right]^{-1} \defo\point
\label{eq:inverse}
\end{equation}

% \AT{the performance of this could be improved by using correctives, leaving here for reference}
% One can also perform a so-called ``reverse'' mapping by instead applying the component-wise inverse for the transforms as defined below:
% 
% \begin{equation}
% \rest\point = \back_\omega(\defo\point \given \pose) = 
% \mathcal{W}_\omega(\defo\point \given \pose) \left[ \T(\pose)^{-1} \right] \defo\point,
% \label{eq:reverse}
% \end{equation}
% This approach does not suffer from the numerical instabilities, but does suffer from severe volume loss.
% Linear blend skinning experiences artifacts known as the ``candy wrapper effect'', and this approach exposes the geometry to the volume loss artifacts along both the forward and reverse mapping operations \cite{posespacedeformation}. 
% Due to the presence of these artefacts, our method uses \eq{inverse} in the inverse mapping.

\subsection{Pose representation}
There are different ways to represent the pose parameters required in input to $\mathcal{W}_\omega$.
One could use the pose parameters defined by an artist and concatenate it to $\defo\point$. 
Alternatively, one can also employ the ``rig-agnostic'' representation suggested by~\cite{nasa} as $(\defo\point \given \pose) {=} \{ \defo\T_b^{-1} \defo\point \}$.
We find this simplifies learning, as the network is directly provided with an encoding of the query in the \textit{local} coordinate frame of each bone.

\begin{figure}[t]
\centering
\begin{overpic} 
[width=\linewidth]
% [width=\linewidth,grid,tics=10]
{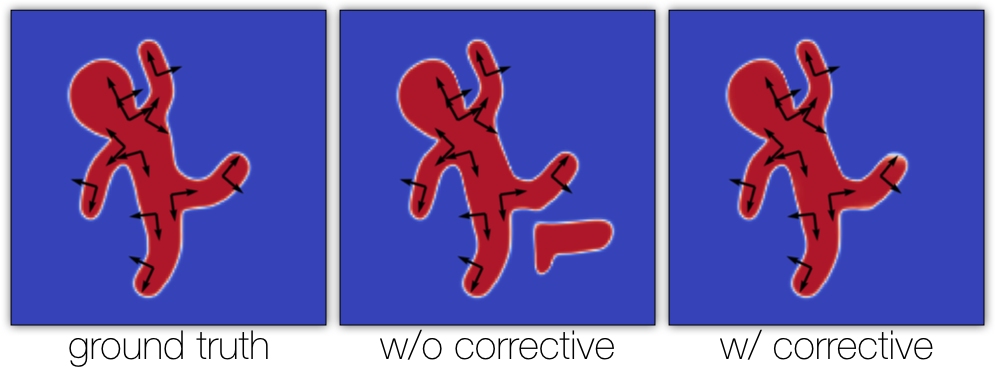}
\end{overpic}
\caption{
\textbf{Corrective --}
The introduction of a \textit{ghost bone} allows us to resolve issues caused by the fact that the softmax has to select at least one bone.
This corrective \textit{ghost bone} allows the network to also reason about the concept of ``background''.
} 
\label{fig:corrective}
\end{figure}
\subsection{Ghost bone corrective}
One can use the inverse mapping~\eq{inverse} to query the occupancy function from the cache:
\begin{equation}
\bar\object(\back_\omega(\defo\point \given \pose)) 
= \bar\object\left( \left[ \mathcal{W}_\omega(\defo\point \given \pose) \right) \T(\pose) \right]^{-1} \defo\point)
\end{equation}
However, due to the fact that the weights in output from $\mathcal{W}_\omega$ satisfy a partition of unity property, the resulting occupancy can contain artefacts;~see \Figure{corrective}~(middle).
To address this, we create an additional ``ghost bone'' modeled by an additional weight network channel, whose transformations are cloned from the root~$\rest\T_{B+1}{=}\rest\T_0$ and $\defo\T_{B+1}{=}\defo\T_0$.
\begin{equation}
    \defo\object_\omega(\defo\point \given \pose) = 
    \left( 1 - \mathcal{W}^{[B+1]}_\omega(\defo\point \given \pose) \right)
    \:\: \bar\object(\back_\omega(\defo\point \given \pose))
\end{equation}
where $\mathcal{W}_\omega(\rest\point){:}\R^3 {\rightarrow} \R^{1{\times}B+1}$. 
As illustrated in \Figure{corrective}, this extra degree of freedom improves results by allowing explicit handling of background regions via the ghost bone.

\subsection{Training}
We train $\omega$ on a ``gingerbread'' dataset consisting of 100 different poses sampled from a temporally coherent animation with poses $\Theta$.
We are only interested in \textit{overfitting} performance, so we have no dataset splits. 
We train by jointly minimizing two losses:
\begin{align*}
\mathcal{L}_\text{occupancy}(\omega) &= \sum_{\pose {\in} \poses}{
\expect{\defo\point \sim \R^3}{\| \defo\object_\omega(\defo\x \given \pose) - \defo\object(\defo\x \given \pose) \|}} 
\\
\loss_{\text{weights}}(\omega) &= \sum_{\vertex_n {\in} \mathbf{V}} \crossentropy{\mathcal{W}_\omega(\rest\vertex_n | \rest\pose), \mathbf{w}_n} 
\end{align*}
where $\crossentropy{.}$ is cross-entropy, and $\loss_{\text{weights}}$ helps to \textit{significantly} speeds up convergence by asking $\mathcal{W}_\omega$ to reproduce LBS deformations on the surface vertices\footnote{Note $\mathbf{w}_n[B+1]=0$ as $\bar{\mathbf{v}}_n$ is on the surface.}.
We employ $\{\mathbf{w}_n\}$ only at \textit{training} time. An example of predicted deformed occupancy is visualized in~\Figure{corrective}.
{
    \small
    \bibliographystyle{plainnat}
    \bibliography{macros,main}
}
\end{document}